\documentclass[runningheads]{llncs}
\usepackage{graphicx}

\usepackage[table]{xcolor}
\usepackage{chngpage}
\usepackage{float}
\usepackage{etoolbox}
\usepackage{amsmath}
\usepackage{rotating}
\usepackage{graphicx}

\usepackage{caption} 
\usepackage{subfig}

\apptocmd{\sloppy}{\hbadness 10000\relax}{}{}

\begin{document}

\title{Automated analysis of diabetic retinopathy using vessel segmentation maps as inductive bias}
\titlerunning{Automated analysis of DR using vessel segmentation maps as inductive bias}

\author{Linus Kreitner\inst{1} \and 
 Ivan Ezhov \inst{1} 
 \and Daniel Rueckert\inst{1,2} \and \\
 Johannes C. Paetzold\inst{2,3\thanks{contributed equally as senior authors}}
 \and Martin J. Menten \inst{1,2 ^*} 
}

\authorrunning{Kreitner et al.}

%
\institute{Lab for AI in Medicine, Klinikum rechts der Isar, Technical University of Munich, Munich, Germany\and
BioMedIA, Department of Computing, Imperial College London, United Kingdom \and
ITERM Institute Helmholtz Zentrum Muenchen, Neuherberg, Germany  \\
\email{linus.kreitner@tum.de}\\ 
}

\maketitle
\setcounter{footnote}{0}
\begin{abstract}
Recent studies suggest that early stages of diabetic retinopathy (DR) can be diagnosed by monitoring vascular changes in the deep vascular complex. In this work, we investigate a novel method for automated DR grading based on ultra-wide optical coherence tomography angiography (UW-OCTA) images. Our work combines OCTA scans with their vessel segmentations, which then serve as inputs to task specific networks for lesion segmentation, image quality assessment and DR grading.
For this, we generate synthetic OCTA images to train a segmentation network that can be directly applied on real OCTA data. We test our approach on MICCAI 2022's DR analysis challenge (DRAC). In our experiments, the proposed method performs equally well as the baseline model.
\vspace{0.6cm}

\end{abstract}

\section{Introduction}
Optical coherence tomography angiography (OCTA) is a non-invasive \textit{in-vivo} method to acquire high-resolution volumetric data from retinal microvasculature. These properties make OCTA efficient for identifying capillary abnormalities, which can often be observed in diabetic retinopathy. DR is a common medical condition that occurs as a result of untreated diabetes mellitus and causes $2.6\%$ of all cases of blindness globally \cite{Sun.2021,Flaxman.2017}. Roughly $11\%$ of all diabetes patients develop vision-threatening complications such as neovascularization (NV), diabetic macular edema, or diabetic macular ischemia. Early stages of DR can be diagnosed, e.g., by non-perfusion areas (NAs) and intraretinal microvascular abnormalities (IRMAs).

A recent survey by Sheng \textit{et al.} calls OCT an “indispensable component of healthcare in ophthalmology”, but notes that there are still many problems to be solved before artificial intelligence (AI) based systems can be reliably integrated in the clinical process \cite{Sheng.2022}. The Diabetic Retinopathy Analysis Challenge (DRAC) is part of MICCAI 2022 in Singapore and aims at advancing machine learning research for DR analysis on OCTA data \cite{BinSheng.2022}. The organizers provide three labeled datasets with more than 1,000 ultra-wide swept-source OCTA (UW-OCTA) images. The samples are $1024\times 1024$ pixel grayscale images and cover different field-of-views (FOVs). The challenge is divided into the tasks A) lesion segmentation, B) image quality assessment, and C) DR grading. Task A) is a pixelwise multi-label segmentation problem for identifying intraretinal microvascular abnormalities, non-perfusion areas, and neovascularization. The public training set contains 109 sample images with 86 cases of IRMA, 106 with NAs, and 35 with NVs. The training set of task B) contains 665 images, $77.9\%$ of which are graded as excellent, $14.6\%$ as good, and $7.5\%$ as poor quality images. Finally, task C) contains 611 samples where $53.7\%$ are classified as healthy, $34.9\%$ as pre-proliferative DR, and the remaining $11.5\%$ as proliferative DR (PDR). The DRAC event follows in the footsteps of similar challenges such as the Diabetic Retinopathy-Grading and Image Quality Estimation Challenge (DeepDRiD) for fundus images \cite{Liu.2022}.

\begin{figure}[htbp]
  \centering
  \includegraphics[width=\columnwidth]{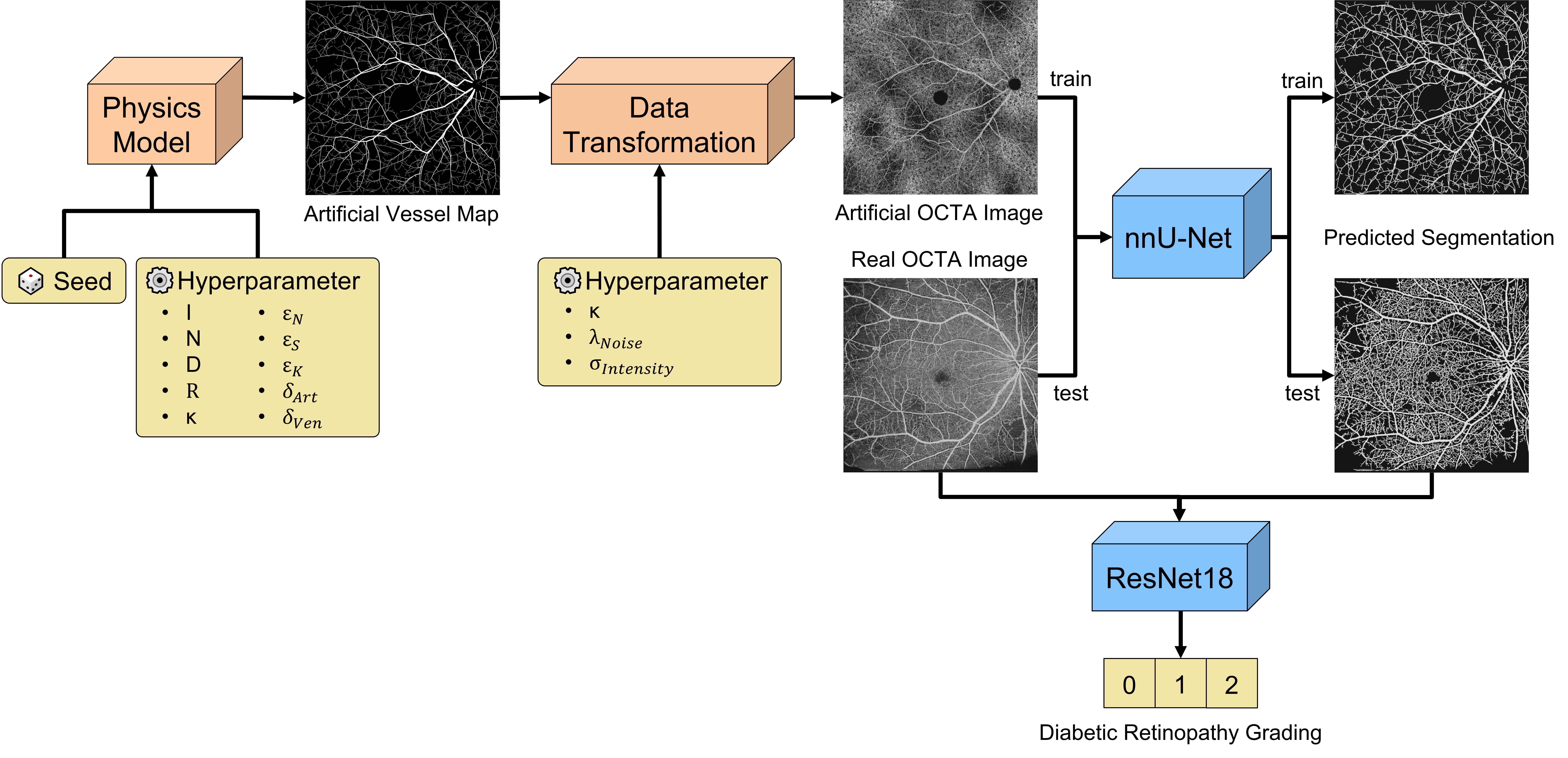}
  \caption{The proposed framework for training a segmentation network using artificial training data. The pre-trained U-Net is then used to generate an additional segmentation layer channel for an arbitrary downstream task.}
  \label{fig:pipeline}
\end{figure}

Diabetic retinopathy causes retinal vascular changes, and while being subtle at first, it eventually leads to dangerous vascular abnormalities and neovascularization. We hypothesize that by explicitly feeding the segmentation map of the vasculature to a deep neural network, we can provide a strong inductive bias for the network to predict the severity of DR. Recent studies suggest that early DR progression is mainly correlated with biomarkers found in the deep vascular complex (DVC), while the superficial vascular complex (SVC) is generally only affected in severe cases \cite{Sun.2021,JacquelineChua.2020}. The authors hypothesize that the DVC might be more susceptible to ischemic damage because of its anatomical proximity to the outer plexiform layer, which has a high oxygen consumption. Therefore, it becomes clear that including small capillaries in the segmentation map is vital. However, to our knowledge, there is no publicly available dataset of OCTA images with corresponding vessel segmentation that includes small capillary vessels. Merely two public datasets, namely OCTA-500 and ROSE, have been released with segmentation labels for vessels and the foveal avascular zone (FAZ), both of which unfortunately only segment the largest vessels of the SVC \cite{Li.14.12.2020,Ma.2021}. To circumvent the problem of missing labeled data, we generate an artificial dataset ourselves. Figure \ref{fig:pipeline} shows our proposed pipeline.

\section{Methodology}
In our work, we investigate whether an additional input channel to our network can boost the performance of a neural network. Our proposed additional input is a segmentation map of the vasculature, including small capillaries in the retina. To extract a faithful segmentation, we use synthetic OCTA images to train a segmentation network, which can then be used to generate the additional input channel on the fly.

\subsection{Synthetic Data Generation}
Synthetic data generation to train segmentation networks is a vastly accepted technique to mitigate data sparsity through transfer learning in medical imaging \cite{raghu2019transfusion,qasim2020red,horvath2022metgan}. 
Schneider \textit{et al.} propose a configurable physics model to simulate the growth of vascular trees based on physical principles \cite{Schneider.2012}. The blood vessel network is represented as a forest of rooted binary tree graphs that are iteratively updated based on the simulated oxygen concentration and the vascular endothelial growth factor (VEGF). Bifurcations and growth direction follow the physical laws of fluid dynamics to ensure realistic branching. From the resulting 3D graph network, it is then possible to retrieve 2D images and their ground truth segmentation maps by voxelizing the edges. This approach has proven efficient across multiple vessel segmentation tasks \cite{gerl2020distance,shit2021cldice,todorov2020machine,paetzold2021whole}. Menten \textit{et al.} later adapted this approach to simulate the retinal vasculature \cite{Menten.2022}. They heuristically tune the simulator's hyperparameters to mimic the structure of the SVC and the DVC. Binary masks are used to decrease the VEGF within the FAZ, as well as the perimeter of the FOV. Using different seeds, it is possible to generate an unlimited amount of diverse vessel maps.

To bridge the domain shift from synthetic to real OCTA images, Menten \textit{et al.} employ a variety of data augmentations, such as eye motion artifacts, flow projection artifacts, and changes in brightness. The background signal caused by small capillary vessels is simulated by binomial noise and convoluted by a Gaussian filter since it is computationally intractable to simulate them. After training a U-Net on a synthetic dataset, the network can extract detailed segmentation maps on real data.

We adopt a modified algorithm version and tune it to realistically represent ultra-wide OCTA images. Instead of setting the roots of the trees on the image's border, we simulate the optical nerve, letting all vessel trees originate from there. 

Instead of limiting bias fields to the outer rings of the FOV, we randomly apply them at every possible position as part of data augmentation. Furthermore, we employ random rotation, flipping, elastic transformation, scaling, and motion artifacts during the training of the segmentation network. We use a variation of the very successful U-Net architecture \cite{Ronneberger.2015} and follow the guidelines of nnU-Net to configure our network optimally for the given task \cite{Isensee.2021}. 

\begin{figure}[htbp]
  \centering
  \subfloat[Synthetic]{ \includegraphics[width=0.31\columnwidth]{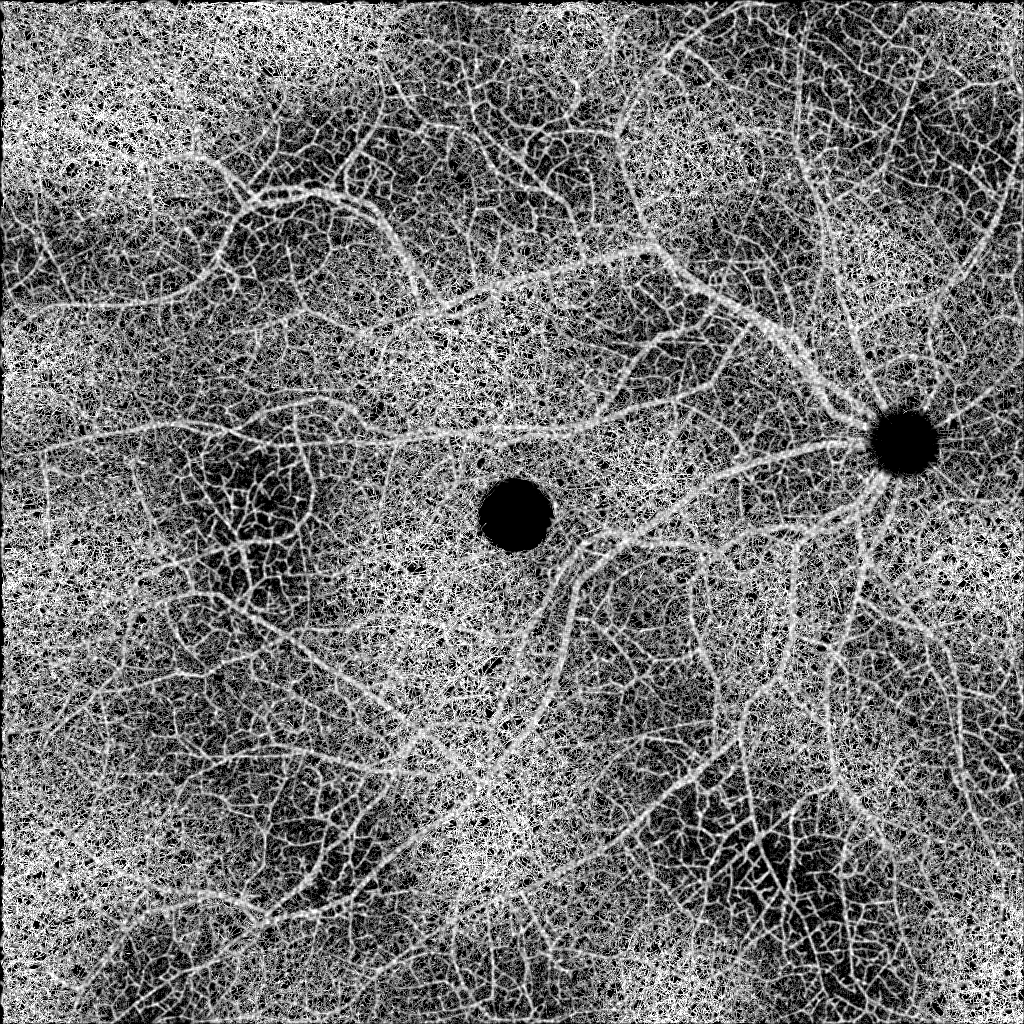} }
  \subfloat[Real]{ \includegraphics[width=0.31\columnwidth]{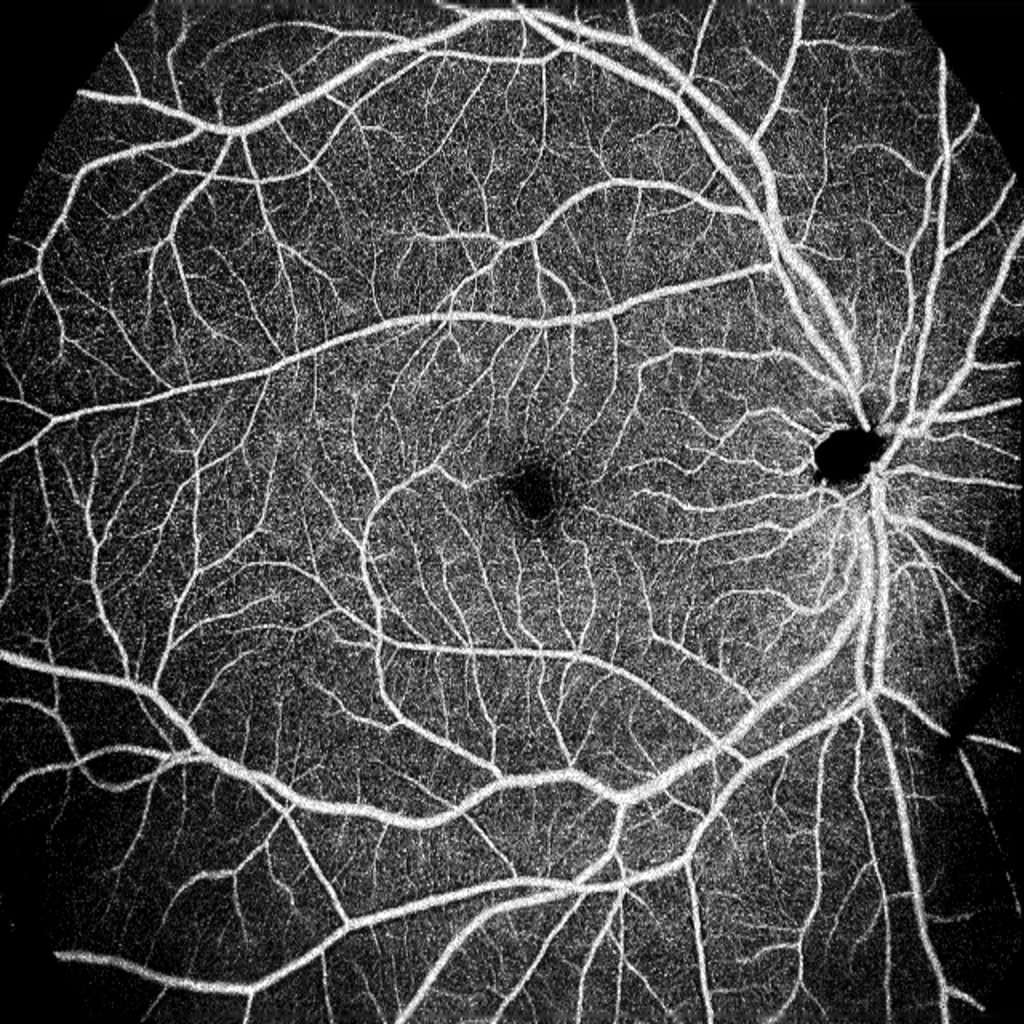} }
  \subfloat[Segmentation]{ \includegraphics[width=0.31\columnwidth]{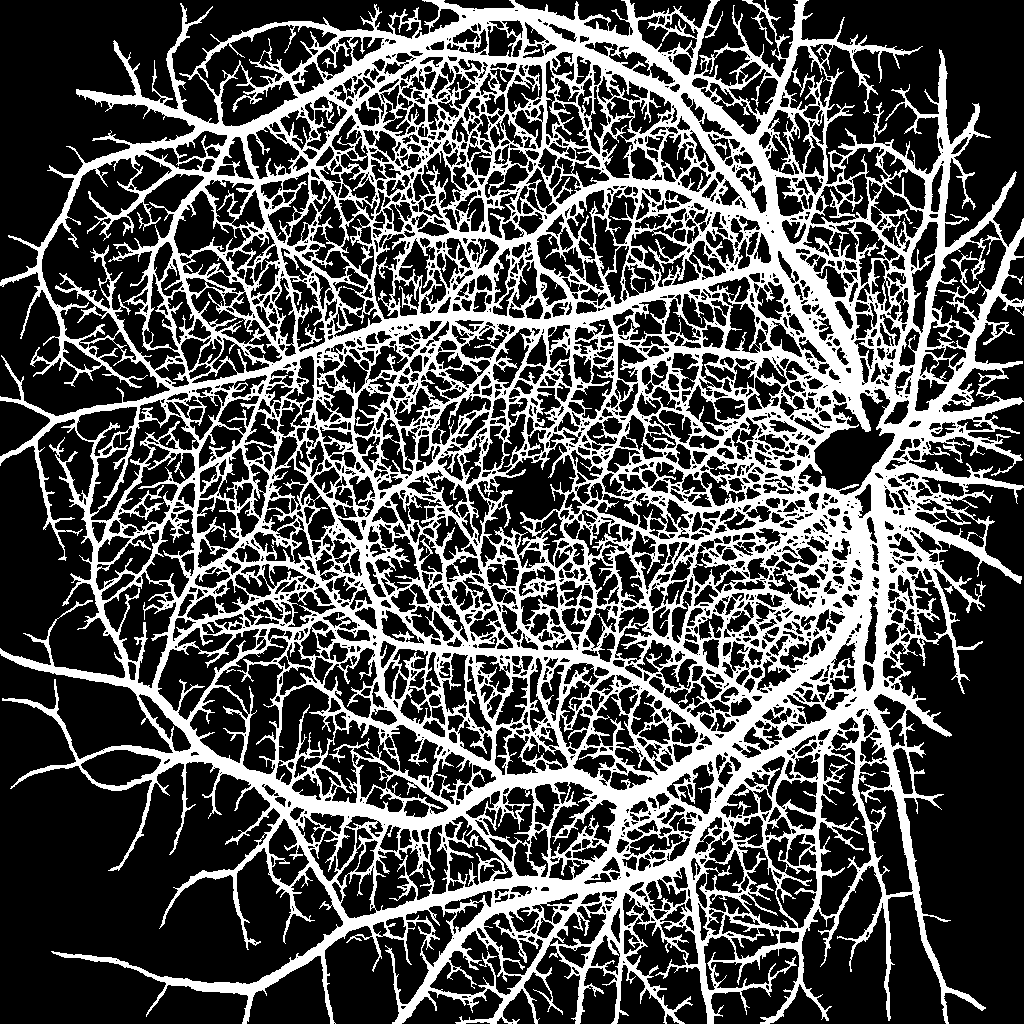} }
  \caption{Comparison of synthetic UW-OCTA images with real samples from the DRAC corpus \cite{BinSheng.2022}. The segmentation map was generated by a network trained only on synthetic data.}
  \label{Synthetic_fig}
\end{figure}

\subsection{Task A: Lesion Segmentation}
\label{section_task1}
\begin{figure}[htbp]
  \centering
  \includegraphics[width=\columnwidth]{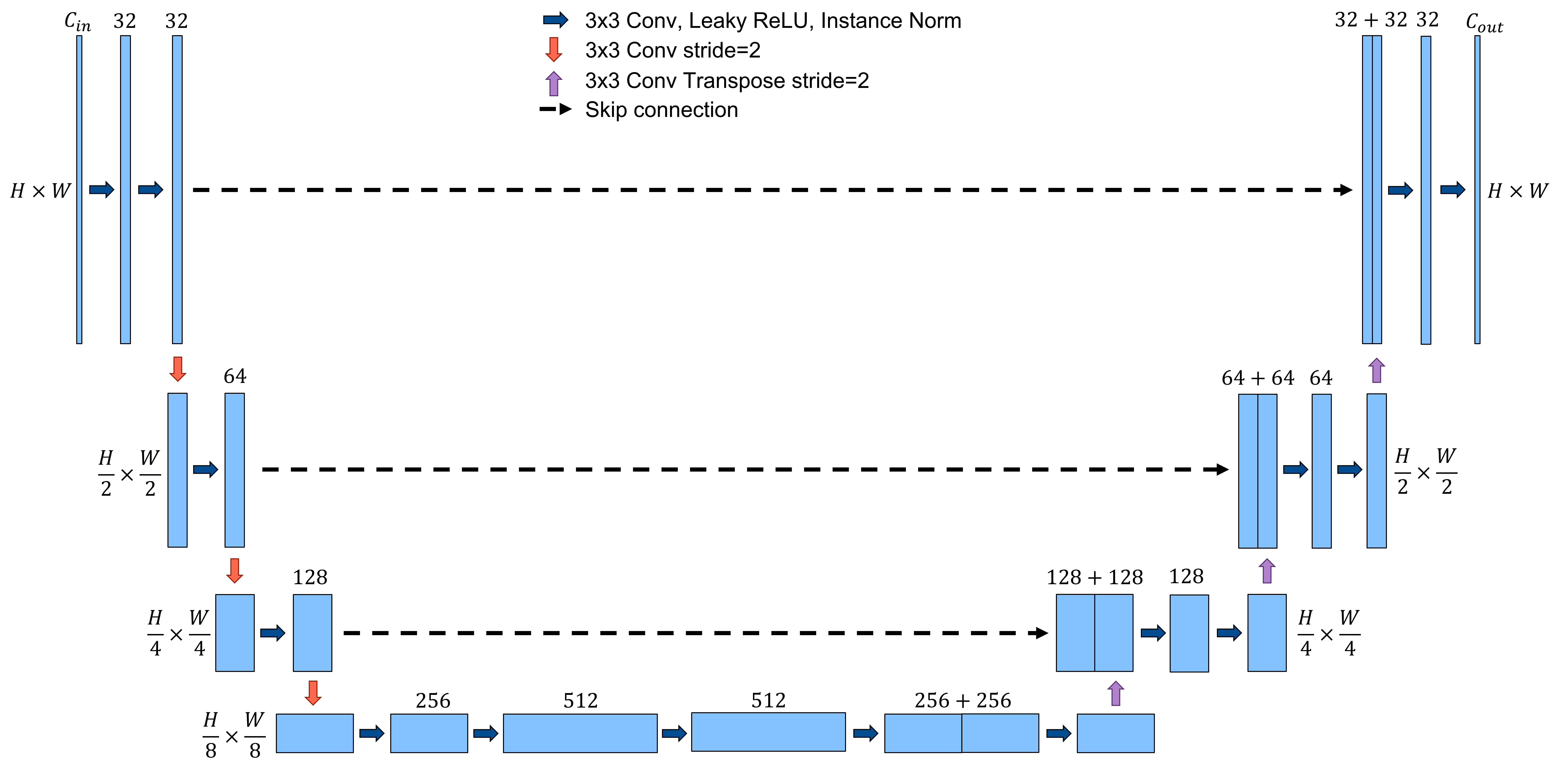}
  \caption{Our U-Net architecture for Task A and for the vessel segmentation.}
  \label{UNet_fig}
\end{figure}

We define the segmentation task as a multi-label classification problem, where each pixel can be part of the classes IRMA, NA, NV, or neither. We choose the widely successful U-Net architecture, which is known for its reliable segmentation performance on medical image datasets \cite{cciccek20163d,Isensee.2021}. For the optimal training parameters, we follow the guidelines of nnU-Net that define preprocessing steps, network architecture, loss function, and learning rate. Our final architecture can be seen in Figure \ref{UNet_fig}, where $C_{in}=1$ for our baseline model, $C_{in}=2$ for the network using the additional segmentation map input, and $C_{out}=3$. Our loss function is the sum of the soft Dice loss and the channel-wise binary cross entropy loss.

Because of the small dataset and large class imbalances, we observe strong overfitting right from the start. We, therefore, apply extensive data augmentation, such as elastic deformation, contrast changes, Gaussian smoothing, flipping, and rotation. To further reduce overfitting, we initialize our proposed network with the weights from the pre-trained vessel segmentation network.
We then train a baseline system as well as our proposed method for 300 epochs and select the checkpoint with the highest Dice similarity coefficient (DSC) on the validation set.

\subsection{Task B: Image Quality Assessment}
Task B is a multi-class classification problem, however, instead of using a classical cross-entropy loss, we redefine the objective as a regression problem using the Mean Squared Error (MSE) loss function. The rationale behind this is that the classes are not independent but rather share a cascading relationship with $C0 \prec C1 \prec C2$. Overfitting is a reoccurring pattern in all tasks. We find that larger models suffer more from overfitting than smaller models, leading us to test two widely known architectures, a ResNet18 and an EfficientNet\_B0 \cite{K.He.2016,Tan.}. Again, we compare the two baseline models with our proposed model, where the vessel segmentation map of an image is added as an additional input. We apply mild elastic deformation, flipping, 90° rotations, and contrast changes, as stronger augmentations may distort the characteristic appearance of the image artifacts. We train all models for 500 epochs and select the model with the best quadratic weighted kappa score.

\subsection{Task C: Diabetic Retinopathy Grading}
Our training method for DR grading is the same as for Task B. We increase the magnitude of the data augmentations and additionally add random rotations by $k\times90^\circ\pm10^\circ$ with $k\in\{0,1,2,3\}$, Gaussian smoothing, and random erasing.

\section{Results}

During the challenge, we only tested the models performing best on a single validation set of each task. As the models all suffer from strong overfitting on the data split and biases by class imbalances, we decided to perform a stratified 5-fold cross-validation in the post challenge phase. We create an ensemble prediction on the test set using all trained models to obtain a more accurate estimation of the real performance. The official challenge test sets contain 65, 438, and 386 samples for tasks A, B, and C, respectively. Even though possible in the DRAC challenge, we specifically did not overfit on the test sets but instead only used the provided training set to select our models. The results for each task are listed in the following.

\subsection{Results for Task A }
To build the ensemble prediction for the segmentation task, we compute the pixel-wise and class-wise average score after applying a sigmoid activation. The output is then transformed into three distinct binary masks with a threshold $t=0.5$. Despite our data augmentations, we observe overfitting. We could not find any significant difference between the baseline and the method using the additional vessel segmentation input. The results are shown in table \ref{table:A}.

\begin{table}[htbp]
\centering
\caption{Results for the lesion segmentation task. We computed the mean Dice Similarity Coefficient (mDSC) and the Intersection of Union (IoU) with their respective standard errors. In general, all methods perform similarly. The highest cross-validation scores are printed in bold.}
\begin{tabular}{|cl|cc|cc|}
    \hline
    \multicolumn{2}{|c|}{Task A} & \multicolumn{2}{c|}{(Cross-)Validation} & \multicolumn{2}{c|}{Challenge Test Set} \\ 
    \cline{3-6} \multicolumn{2}{|c|}{} & \multicolumn{1}{c|}{mDSC} & IoU & \multicolumn{1}{c|}{mDSC} & IoU \\
    \hline
    \multicolumn{1}{|c|}{nnU-Net} & Base+Seg &
    \multicolumn{1}{c|}{0.086} &
    {0.459} &
    \multicolumn{1}{c|}{0.41} &
    {0.284} \\
    \hline
    \multicolumn{1}{|c|}{nnU-Net} & Base &
    \multicolumn{1}{c|}{\textbf{0.558\textpm0.015}} &
    {\textbf{0.424\textpm0.015}} &
    \multicolumn{1}{c|}{\textbf{0.529\textpm0.011}} &
    \textbf{0.385\textpm0.011} \\ 
    
    \cline{2-6} \multicolumn{1}{|c|}{Ensemble} & Base+Seg &
    \multicolumn{1}{c|}{0.556\textpm0.018} &
    0.422\textpm0.018 &
    \multicolumn{1}{c|}{0.518\textpm0.014} &
    0.377\textpm0.013 \\
    \hline
\end{tabular}
\label{table:A}
\end{table}

\subsection{Results for Task B }
For the image classification task and the DR grading, the soft outputs from all trained models are averaged to form the ensemble prediction. First, comparing both architectures, we cannot find a significant performance gain from one over the other. During training, however, the EfficientNet was much more stable in its predictions, while the ResNet showed large jumps. Furthermore, the additional segmentation input did not seem to have meaningful benefits for training but instead caused the model to overfit faster. All results are listed in Table \ref{table:B}.

\begin{table}[htbp]
\centering
\caption{Results for the image quality assessment task. We computed the Quadratic Weighted Kappa (QWK) score as well as the Area Under receiver operating characteristic Curve (AUC) with their respective standard errors. All methods perform similarly, with the EfficientNet being the most stable to train. The highest cross-validation scores are printed in bold.}
\begin{tabular}{|cl|cc|cc|}
    \hline
    \multicolumn{2}{|c|}{Task B} & \multicolumn{2}{c|}{(Cross-)Validation} & \multicolumn{2}{c|}{Challenge Test Set} \\
    \cline{3-6} \multicolumn{2}{|c|}{} & \multicolumn{1}{c|}{QWK} & AUC & \multicolumn{1}{c|}{\ \ QWK\ \ \ } & {AUC} \\
    \hline
    \multicolumn{1}{|c|}{ResNet18} & Base+Seg &
    \multicolumn{1}{c|}{0.917} & 0.931 &
    \multicolumn{1}{c|}{0.677} & {0.802} \\
    \hline
    \multicolumn{1}{|c|}{ResNet18} & Base &
    \multicolumn{1}{c|}{0.891\textpm0.011} &
    \textbf{0.956\textpm0.010} &
    \multicolumn{1}{c|}{\textbf{0.726}} &
    0.831 \\ 
    \cline{2-6} \multicolumn{1}{|c|}{Ensemble} & Base+Seg &
    \multicolumn{1}{c|}{0.891\textpm0.011} & 0.932\textpm0.009 &
    \multicolumn{1}{c|}{0.699} & \textbf{0.843} \\
    \hline
    \multicolumn{1}{|c|}{EfficientNet\_B0} & Base  &
    \multicolumn{1}{c|}{\textbf{0.901\textpm0.010}} &
    0.930\textpm0.010 & \multicolumn{1}{c|}{0.707} &
    0.837 \\
    \cline{2-6}  \multicolumn{1}{|c|}{Ensemble} & Base+Seg  &
    \multicolumn{1}{c|}{0.893\textpm0.012} &
    0.953\textpm0.010 & 
    \multicolumn{1}{c|}{0.662} & 0.818 \\
    \hline
\end{tabular}
\label{table:B}
\end{table}

\subsection{Results for Task C }
In the DR grading task, we employ the same approach as in Task B to compute the ensemble prediction. Again, we could not find a significant difference between the EfficientNet and the ResNet architecture, but the smoother training for the EfficientNet is also observed here. While the additional segmentation input did not yield any measurable benefit during cross-validation, the EfficientNet using the additional input outperformed the other models on the test set by a substantial margin. It is difficult to interpret, however, if this was caused by the additional information or due to random chance causing one version to perform better on the test set. Table \ref{table:C} depicts all results in detail.

\begin{table}[htbp]
\caption{Results for the diabetic retinopathy grading task. We computed the Quadratic Weighted Kappa score and the Area Under the receiver operating characteristic Curve with their respective standard errors. The EfficientNet seems to outperform the ResNet by a small margin but no substantial advantage for the additional segmentation input can be observed. The highest cross-validation scores are printed in bold.}
\centering
\begin{tabular}{|cl|ll|ll|}
\hline
    \multicolumn{2}{|c|}{Task C} & \multicolumn{2}{c|}{(Cross-)Validation} & \multicolumn{2}{c|}{Challenge Test Set} \\
    \cline{3-6}  \multicolumn{2}{|c|}{} & \multicolumn{1}{c|}{QWK} & \multicolumn{1}{c|}{AUC} & \multicolumn{1}{c|}{\ \ QWK\ \ \ } & \multicolumn{1}{c|}{AUC} \\
    \hline
    \multicolumn{1}{|c|}{Resnet18} & Base+Seg &
    \multicolumn{1}{c|}{0.838} &
    \multicolumn{1}{c|}{0.894} &
    \multicolumn{1}{c|}{0.277} &
    \multicolumn{1}{c|}{0.666} \\
    \hline
    \multicolumn{1}{|c|}{Resnet18} & Base &
    \multicolumn{1}{c|}{0.831\textpm0.014} &
    \multicolumn{1}{c|}{0.911\textpm0.010} &
    \multicolumn{1}{c|}{0.824} &
    \multicolumn{1}{c|}{0.891} \\
    \cline{2-6} \multicolumn{1}{|c|}{Ensemble} & Base+Seg &
    \multicolumn{1}{c|}{0.846\textpm0.014} &
    \multicolumn{1}{c|}{0.922\textpm0.010} &
    \multicolumn{1}{c|}{0.805} &
    \multicolumn{1}{c|}{0.891} \\
    \hline
    \multicolumn{1}{|c|}{EfficientNet\_B0} & Base &
    \multicolumn{1}{c|}{0.845\textpm0.010} &
    \textbf{0.929\textpm0.006} &
    \multicolumn{1}{c|}{0.811} &
    \multicolumn{1}{c|}{0.889} \\
    \cline{2-6} \multicolumn{1}{|c|}{Ensemble} & Base+Seg &
    \multicolumn{1}{c|}{\textbf{0.850\textpm0.009}} &
    \multicolumn{1}{c|}{0.918\textpm0.004} &
    \multicolumn{1}{c|}{\textbf{0.843}} &
    \multicolumn{1}{c|}{\textbf{0.909}} \\
    \hline
\end{tabular}
\label{table:C}
\end{table}

\section{Discussion}
This study has investigated whether an additional channel containing the vessel segmentation map can boost the performance of a baseline model for the tasks of lesion segmentation, image quality assessment, and DR grading. This was motivated by the observation that early stages of DR manifest themselves in vascular changes of the DVC. We hypothesized that explicitly adding a vessel map as inductive bias would reduce overfitting through smaller segmentation and classification networks. While we were not able to win the DRAC challenge, we point out that this was not the primary goal of this study. Every team member was allowed to evaluate a model on the test set once per day, enabling teams with multiple members to test repeatedly and select their best model based on the test set. Additionally, the test set images were made public, making it possible to label the images manually. However, we deliberately abstained from extensive model tuning and overfitting on the test set.
 
 Instead, we focused on testing the idea of integrating synthetic data into the prediction life cycle. We successfully extended Menten \textit{et al.}'s work to generate OCTA images for ultra-wide scans. Using our training method, we can segment tiny capillaries whose diameter is similar to the physical resolution of the OCTA scanner. On this dataset, we could not find any significant improvement using the segmentation maps as input. However, we consider this finding relevant, as it indicates that the baseline models are strong enough to extract that information by themselves. It would be interesting to see whether this method could benefit from a larger image corpus, where networks are less prone to overfitting.
 
 For future work, we aim to use the generated segmentation map to extract quantitative biomarkers. These in turn could be used to, e.g., train a random forest for DR grading. The prediction would hence become more explainable and allow doctors to retrace the network's reasoning.

\section{Conclusion}
This paper summarizes our contribution to the DR analysis challenge 2022. We present a novel method, where synthetic OCTA images are used to train a segmentation network, which then generates an additional input channel for downstream tasks. Our method extracts state-of-the-art segmentation maps including tiny vessels from the deep vascular complex. However, we could not measure a substantial benefit of adding this auxiliary information to our deep learning pipeline in any of the three challenge tasks. Future work might build on our idea to use segmentation maps as complementary information and explore other applications where this is beneficial.


{ \bibliographystyle{splncs}
\bibliography{main}
}

\end{document}